\documentclass{elsart}

\usepackage{graphicx}
\usepackage{epsfig}

\def\Journal#1#2#3#4{{#1} {#2} (#4) #3}
\def\NIMR{Nucl. Instr. and Meth. A}
\def\NPA{Nucl. Phys. A}
\def\PLB{Phys. Lett.  B}
\def\PRL{Phys. Rev. Lett.}
\def\PRC{Phys. Rev. C}
\def\PRD{Phys. Rev. D}
\def\PR{Phys. Rep.} 
\def\ZPA{Z. Phys. A}

\def\ARNPS{Ann. Rev. Nucl. Part. Sc.}

\def\PPNP{Prog. Part. Nucl. Phys.}
\def\be{\begin{equation}}
\def\ee{\end{equation}}

\newcommand{\pt}{$p_t^{(0)}$}

\newcommand{\avpt}{$\langle p_t^{(0)} \rangle$}

\begin{document}
\begin{frontmatter}

\title{Excitation function of elliptic flow in Au+Au collisions and the 
nuclear matter equation of state} 

\vspace{0.5cm}
\author[gsi]{A.\,Andronic\thanksref{info}},
\author[clt]{V.\,Barret}, 
\author[zag]{Z.\,Basrak}, 
\author[clt]{N.\,Bastid}, 
\author[hei]{L.\,Benabderrahmane}, 
\author[bud]{G.\,Berek},
\author[zag]{R.\,\v{C}aplar},
\author[hei]{E.\,Cordier}, 
\author[clt]{P.\,Crochet}, 
\author[clt]{P.\,Dupieux}, 
\author[zag]{M.\,D\v zelalija},
\author[bud]{Z.\,Fodor},
\author[zag]{I.\,Gasparic}, 
\author[ite]{Yu.\,Grishkin},
\author[gsi]{O.N.\,Hartmann}, 
\author[hei]{N.\,Herrmann}, 
\author[gsi]{K.D.\,Hildenbrand}, 
\author[kor]{B.\,Hong},
\author[bud]{J.\,Kecskemeti}, 
\author[kor]{Y.J.\,Kim},
\author[gsi,hei,war]{M.\,Kirejczyk}, 
\author[gsi]{P.\,Koczon}, 
\author[zag]{M.\,Korolija}, 
\author[dre]{R.\,Kotte},
\author[gsi]{T.\,Kress},
\author[ite]{A.\,Lebedev}, 
\author[gsi]{Y.\,Leifels}, 
\author[clt]{X.\,Lopez}, 
\author[hei]{A.\,Mangiarotti},
\author[hei]{M.\,Merschmeyer},
\author[dre]{W.\,Neubert},
\author[hei]{D.\,Pelte}, 
\author[buc]{M.\,Petrovici},
\author[ire]{F.\,Rami},
\author[gsi]{W.\,Reisdorf},
\author[ire]{B.\,de Schauenburg}, 
\author[gsi]{A.\,Sch\"uttauf},
\author[bud]{Z.\,Seres},
\author[war]{B.\,Sikora}, 
\author[kor]{K.S.\,Sim}, 
\author[buc]{V.\,Simion}, 
\author[war]{K.\,Siwek-Wilczy\'nska}, 
\author[ite]{V.\,Smolyankin}, 
\author[hei]{M.R.\,Stockmeier}, 
\author[buc]{G.\,Stoicea}, 
\author[gsi,war]{Z.\,Tyminski},
\author[ire]{P.\,Wagner}, 
\author[hei,war]{K.\,Wi\'sniewski}, 
\author[dre]{D.\,Wohlfarth},
\author[gsi]{Z.-G.\,Xiao},
\author[kur]{I.\,Yushmanov},
\author[ite]{A.\,Zhilin} 

\vspace{0.25cm}
\address[buc]{National Institute for Physics and Nuclear Engineering, Bucharest, Romania}
\address[bud]{KFKI Research Institute for Particle and Nuclear Physics, Budapest, Hungary}
\address[clt]{Laboratoire de Physique Corpusculaire, IN2P3/CNRS,
and Universit\'{e} Blaise Pascal, Clermont-Ferrand, France}
\address[gsi]{Gesellschaft f\"ur Schwerionenforschung, Darmstadt, Germany}
\address[dre]{IKH, Forschungszentrum Rossendorf, Dresden, Germany}
\address[hei]{Physikalisches Institut der Universit\"at Heidelberg, Heidelberg, Germany}
\address[ite]{Institute for Theoretical and Experimental Physics, Moscow, Russia}
\address[kur]{Kurchatov Institute, Moscow, Russia}
\address[kor]{Korea University, Seoul, South Korea}
\address[ire]{Institut de Recherches Subatomiques, IN2P3-CNRS, Universit\'e
Louis Pasteur, Strasbourg, France}
\address[war]{Institute of Experimental Physics, Warsaw University, Poland}
\address[zag]{Rudjer Boskovic Institute, Zagreb, Croatia}

(FOPI Collaboration)
 
\thanks[info]{Corresponding author: GSI, Planckstr. 1, 64291 Darmstadt,
Germany; Email:~A.Andronic@gsi.de; Phone: +49 615971 2769; 
Fax: +49 615971 2989.}

\begin{abstract}

We present measurements of the excitation function of elliptic flow 
at midrapidity in Au+Au collisions at beam energies from 0.09 to 1.49 GeV 
per nucleon.
For the integral flow, we discuss the interplay between collective 
expansion and spectator shadowing for three centrality classes.
A complete excitation function of transverse momentum dependence 
of elliptic flow is presented for the first time in this energy range,
revealing a rapid change with incident energy below 0.4~$A$GeV,
followed by an almost perfect scaling at the higher energies.
The equation of state of compressed nuclear matter is addressed 
through comparisons to microscopic transport model calculations.

\end{abstract}


\end{frontmatter}

The study of collective flow in nucleus-nucleus collisions has been 
an intense field of research for the past twenty years \cite{rev,her}. 
At beam energies below several GeV per nucleon, the main motivation 
for studying flow is the extraction of the equation of state (EoS) of 
nuclear matter.  
This can only be accomplished via comparisons to theoretical transport 
models which treat the collision at a microscopic level \cite{sto86,dan}.
It is essentially due to the competing effects of two-body collisions 
and mean field dependences that no firm conclusion on EoS is established 
for the moment \cite{dan}.

Elliptic flow at midrapidity (called "squeeze-out" in the early days)
is a prominent collective flow observable that has received great attention 
in the past.
After the pioneering measurements at Saturne \cite{dem} and Bevalac \cite{gut},
a wealth of experimental results have been obtained at Bevalac and SIS 
\cite{wan,bri,tsa,bas,cro1,and1,sto,indra} as well as at AGS \cite{pin,pbm},
SPS \cite{cer,na49}, and RHIC \cite{star}.
Elliptic flow is consequently established as a powerful observable in the
study of relativistic nucleus-nucleus collisions \cite{oll,vol}.
In what concerns the range of beam energies of 0.1-10~$A$GeV, recent 
microscopic transport model calculations \cite{dan,dan2,lar,per,gai}
have emphasized the importance of elliptic flow for imposing constraints
on the models, towards the extraction of EoS.
In this energy range, the densities, extracted from transport model 
calculations \cite{dan}, are up to several times normal nuclear matter 
density. 
Probing EoS at such densities has implications to astrophysical
questions \cite{dan}.
At beam energies below 2~$A$GeV, no existing set of experimental data 
can be compared on equal footing to the measurements at higher energies. 
Moreover, recent experimental results have demonstrated that directed flow 
is correlated with transparency \cite{rei}.
To complete the picture, it is important to correlate also elliptic flow
with these observations.

In this paper we present new elliptic flow data at midrapidity for Au+Au
collisions for the energy range 0.09-1.49~$A$GeV.
The data have been measured with an almost complete phase-space coverage 
using the FOPI detector \cite{fopi} at SIS, GSI. 
The reaction products are identified according to their charge ($Z$) in 
the forward Plastic Wall (PW) at 1.2$^\circ <\theta_{lab}<$~30$^\circ$ 
using time-of-flight (ToF) and specific energy loss. 
In the Central Drift Chamber (CDC), covering 
34$^\circ <\theta_{lab}<$~145$^\circ$, the particle identification is 
on mass ($A$), obtained using magnetic rigidity and specific 
energy loss. 
For the beam energies above 0.4~$A$GeV, measured in a separate run,
the forward drift chamber Helitron, covering the interval 
7$^\circ <\theta_{lab}<$~29$^\circ$ is employed for particle identification 
on $A$.
We use normalized center-of-mass (c.m.) transverse momentum (per nucleon) 
and rapidity 
$p_t^{(0)}=(p_t/A)/(p_P^{\mathrm{c.m.}}/A_P), 
\quad 
y^{(0)}=(y/y_P)^{\mathrm{c.m.}},
\label{eq-1}
$
where the subscript $P$ denotes the projectile.
Our midrapidity selection is $|y^{(0)}|<$~0.1.
Most of our results are for $Z$=1 particles.
For the beam energies above 0.4~$A$GeV we also present integral elliptic flow
for protons.
Identified by CDC and Helitron, pions are excluded from the $Z$=1 sample. 
For the acceptance of CDC ($p_t^{(0)}>0.8$, at midrapidity) $^3$He is included
in the $Z$=1 sample. 
For the centrality selection we use the charged particle multiplicities, 
classified into five bins, M1 to M5. 
We present results for the centrality bins M2, M3, and  M4, which
correspond on average to the geometric impact parameter intervals 
7.5-9.5 fm, 5.5-7.5 fm, and 2-5.5 fm, respectively. 
Variations of these ranges for different energies, are up to 5\% for the 
energies above 0.4~$A$GeV and up to 8\% for the other energies.

The reaction plane is reconstructed event by event using the transverse
momentum method \cite{dan85}.
All charged particles detected in an event are used, except a window around 
midrapidity ($|y^{(0)}|<$0.3) to improve the resolution.
The correction of the extracted $v_2$ values due to the reconstructed 
reaction plane fluctuations is done using the recipe of Ollitrault \cite{oli}. 
The correction factor is energy- and centrality-dependent.
For the M3 and M4 centralities this factor is as low as 1.2 for energies 
around 0.4~$A$GeV, reaches 1.6 at 1.49~$A$GeV and is between 2 and 4.5 for 
the energies of 0.12 and 0.09~$A$GeV.
For the centrality M2, the correction factor is about 1.6 for energies 
around 0.4~$A$GeV, reaches 2.8 at 1.49~$A$GeV and 4 at 0.12~$A$GeV.

Elliptic flow is quantified by the second order Fourier coefficient 
$v_2=\langle\cos(2\phi)\rangle$, where $\phi$ is the angle with respect 
to the reaction plane.
The systematic errors are for most of our energies of the order of 10\% 
for M2 and M3 bins and slightly larger for M4. 
They increase with the beam energy, reaching at 1.49~$A$GeV about 20\%.
Except otherwise stated, in the following plots, the errors are the 
systematic ones. 
They have three sources:
i) the interpolation over the region $\theta_{lab}=30^\circ - 34^\circ$,
corresponding to \pt=0.6-0.8 at midrapidity, which is not covered by 
FOPI detector.
ii) pion contamination in the $Z$=1 sample, as for \pt$<$0.2 the pions are 
rejected on a statistical basis only. 
iii) particle losses due to multiple hits, which occur preferentially in 
the reaction plane due to the strong directed flow \cite{and2}. 

\begin{figure}[htb]
\centering\mbox{\epsfig{file=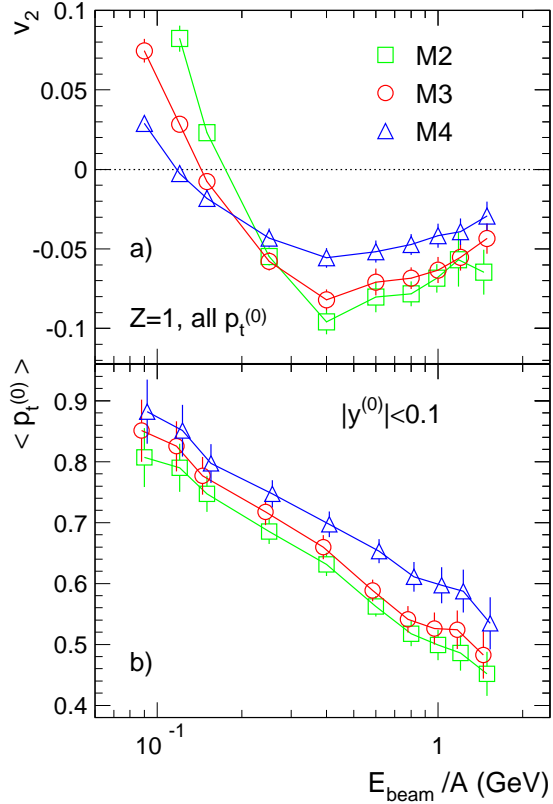, width=.48\textwidth}}
\caption{Excitation function for three centrality bins for:
a) elliptic flow integrated on momentum;
b) average normalized transverse momentum.
}
\label{fig-0}
\end{figure} 

In Fig.~\ref{fig-0} we present, for $Z$=1 particles, the excitation function
of elliptic flow integrated over \pt~ and of the average values of normalized
transverse momentum, \avpt, for three centrality bins.
The elliptic flow evolves from a preferential in-plane, rotational-like,
emission ($v_2>0$) to an out-of-plane, or "squeeze-out" ($v_2<0$) pattern,
as the energy increases.
As seen in Fig.~\ref{fig-0}a, this transition \cite{bas,and1} depends on 
centrality.
Beyond the transition, for higher energies, the strength of the collective 
expansion overcomes the rotational-like motion \cite{cro1}. 
This leads to an increase of out-of-plane emission towards a maximum at 
0.4~$A$GeV.
Beyond 0.4~$A$GeV, elliptic flow decreases towards a transition to in-plane 
preferential emission \cite{pin}. 
In our energy range, the energy dependence of elliptic flow is very similar
to that of directed flow \cite{rev,her,rei}.
As a consequence of comparable passing times and expansion times, in this
energy regime, elliptic flow results as an interplay between fireball
expansion and spectator shadowing.
At sufficiently high incident energy, out-of-plane emission is expected
to grow with: i) the achieved pressure in the fireball (leading to larger 
\avpt~ values, as seen in  Fig.~\ref{fig-0}b), which is larger
for more central collisions ; ii) the degree of shadowing, which is 
proportional with the impact parameter.
This second behavior is observed in the data for energies of 0.4~$A$GeV and 
above, Fig.~\ref{fig-0}a.
We wish to stress that this phenomenon is quite different from what is
known at much higher energies, where higher impact parameters lead to 
increased in-plane emission \cite{oll,vol,star,na49}.
In this case, corresponding to a much shorter passing time compared to 
the expansion time, shadowing by still incoming participants and by spectators
takes place only in the very early stages of the collision. 
The later, dominant, stages of flow generation are determined by the almond 
shape of the isolated fireball \cite{oll}.

In our energy domain, due to increasing transparency in smaller fireballs
\cite{rei}, the growing of the out-of-plane emission towards more peripheral 
collisions will be partially quenched.
To elucidate this, it is useful to gather information on generated transverse
momenta.
As seen in Fig.~\ref{fig-0}b, the scaled transverse momenta, \avpt~ decrease
both as a function of centrality and as a function of incident energy. 
In the low energy end of the data, some of the effect as a function of energy
is likely due to the decreasing importance, in scaled units, of Fermi motion,
while at the high energy end, increasing transparency is suggested by stopping
systematics \cite{rei}. 
In this domain, it is the decrease of \avpt~ which mainly determines the 
decrease of $v_2$ as a function of energy (see next Figure).

\begin{figure}[htb]
\centering\mbox{\epsfig{file=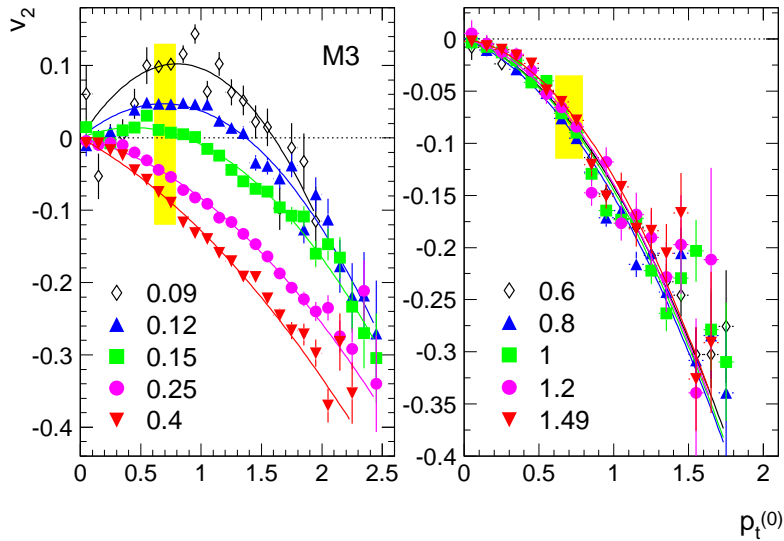, width=.68\textwidth}}
\caption{Excitation function of differential elliptic flow for $Z$=1 particles,
for M3 centrality.
The lower and higher energies are shown separately in the left and right 
panels, respectively.
The symbols represent the measurements (the errors are statistical) and 
the lines are parabolic fits.
In the region not covered by the detector (\pt=0.6-0.8, shades)
the symbols are the result of the fit. 
} 
\label{fig-1}
\end{figure} 

More details on the energy dependence of elliptic flow are revealed by
a study of the transverse momentum dependence of $v_2$, the so-called 
differential flow.
In Fig.~\ref{fig-1} we show for $Z$=1 particles the differential $v_2$
for all incident energies, for centrality bin M3.
In the low energy region, differential elliptic flow exhibits a dramatic
change as a function of beam energy. It evolves from a coexistence
of in-plane and out-of-plane features towards a monotonic (parabolic) 
dependence as a function of \pt. 
This change may be a result of a gradual transition of the collision regime 
from a mean field to a nucleon-nucleon scattering domination.
As seen in Fig.~\ref{fig-1}, the transition from in-plane to out-of-plane
depends on \pt, a feature which we have studied earlier in detail \cite{and1}.
This is markedly different than in case of the transition (at about 4~$A$GeV)
from out-of-plane to in-plane, which was found to be independent on $p_t$ 
\cite{pin}.
For energies beyond 0.4~$A$GeV, our data show that elliptic flow approximately
scales as a function of \pt, a feature noted earlier for low-\pt~ FOPI data 
\cite{bas}. 
In view of this scaling, it is the decreasing \avpt~ as a function of energy 
(Fig.~\ref{fig-0}b) which explains the decrease of the absolute magnitude
of the integral elliptic flow for energies above 0.4~$A$GeV.
The scaling of differential $v_2$ is also supported by proton differential
elliptic flow measured by KaoS collaboration \cite{bri} and was noted 
as well in the case of pions \cite{rev}.
We interpret it as a consequence of an approximate matching between
the time scale of the expansion and of the passing time of the spectators,
$t_{pass}$. In a simple geometric participant-spectator model, 
$t_{pass}=2R/(\gamma_s v_s)$, where $R$ is the radius of the nucleus
at rest, $v_s$ is the spectator velocity in c.m. and $\gamma_s$ the 
corresponding Lorentz factor.
In the energy range 0.4-1.49~$A$GeV, $t_{pass}$ decreases from 30 to 16 fm/c,
implying that overall the expansion gets about two times faster in this 
energy range. This is also supported by the average expansion velocities
extracted from particle spectra \cite{wan}.
High-\pt~ particles, for which elliptic flow is larger, are emitted at 
times even shorter than $t_{pass}$.
Originating from the largest density region, they probe the momentum dependent
interactions of the mean field in a unique way \cite{dan2}.
In a recent paper \cite{sto} we have shown that detailed time-like 
diagnosis on the expansion dynamics can be obtained from the azimuthal 
dependence of the collective flow energy.
We note that emission times of 13 fm/c were extracted for high-energy pions
at spectator rapidity in Au+Au collisions at 1~$A$GeV 
\cite{wag}.

\begin{figure}[htb]
\centering\mbox{\epsfig{file=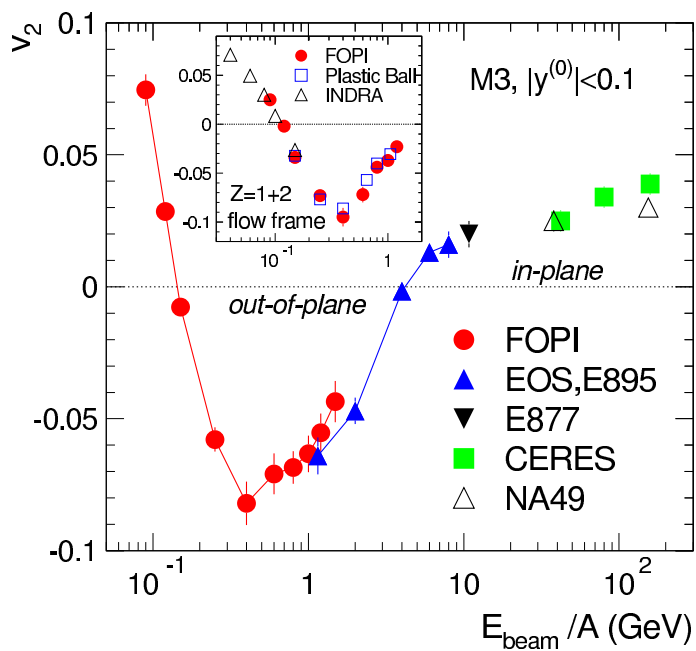, width=.57\textwidth}}
\caption{Excitation function of integral elliptic flow for the M3 
centrality bin. The FOPI data for $Z$=1 particles are compared to measurements 
available at all fixed target experiments: EOS and E895 \cite{pin} for protons,
E877 \cite{pbm} and CERES \cite{cer} for all charged particles
and NA49 \cite{na49} for pions.
The inset shows $v_2$ data for $Z\le$2 particles, analyzed in the directed 
flow reference frame (see text).
} 
\label{fig-2}
\end{figure} 

An excitation function of the integral elliptic flow is presented 
in Fig.~\ref{fig-2} for all existing measurements 
at fixed-target experiments.
Our data for $Z$=1 particles are compared to the data of experiments
EOS and E895 \cite{pin} for protons,
E877 \cite{pbm} and CERES \cite{cer} for all charged particles
and NA49 \cite{na49} for pions.
The centrality classes for the various experiments are similar to our 
M3 centrality bin, which corresponds to the range 15\%-29\% of the 
geometrical cross section.
The composite particles included in our $Z$=1 sample are important only 
for energies below 0.8~$A$GeV, while above, due to their low yield, they
do not contribute anymore to enhance the $Z$=1 flow with respect to proton
(see next Figure).
The increase of elliptic flow beyond 10~$A$GeV, which is logarithmic in 
$\sqrt{s_{NN}}$ up to RHIC energies \cite{aa}, is determined by the 
gradient-driven expansion of the unhindered almond-shape fireball \cite{oll}.

The inset in Fig.~\ref{fig-2} shows an excitation function of $v_2$
in the reference frame with the $z$ axis (which is usually the beam direction)
along the main axis of the directed flow. 
To allow for a direct comparison to the results of Plastic Ball experiment 
at Bevalac \cite{gut}, the results are for $Z\le$2 particles, integrated on 
\pt.
Our data is in a very good agreement with the Bevalac results 
and also agree very well with recent data measured by Indra collaboration 
\cite{indra}. The Indra data are for the impact parameter range 4.5-6.5 fm,
which is slightly more central than the FOPI and Plastic Wall centrality
bin.
Note that, as the Bevalac data were not corrected for the reaction plane 
resolution, no correction is employed for the FOPI and Indra data either.
Since all have a quasi-complete coverage of the phase space, the correction
is similar for the three experiments.
It influences the shape of the excitation function, in particular at low 
and high energies.

\begin{figure}[htb]
\vspace{-.8cm}
\centering\mbox{\epsfig{file=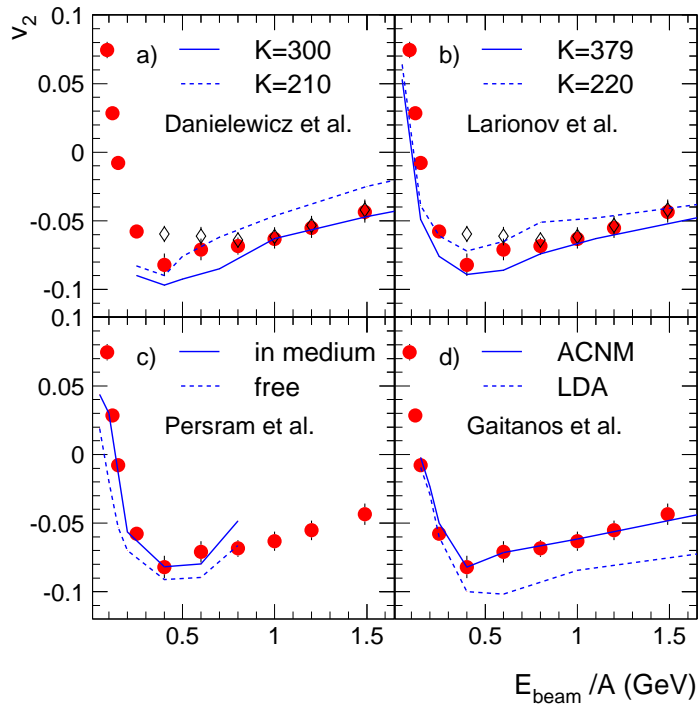, width=.67\textwidth}}
\caption{{Measured elliptic flow for $Z$=1 (dots) and protons (diamonds) 
for M3 centrality in comparison to BUU model calculations (lines) of:
a) Danielewicz et al. \cite{dan} and b) Larionov et al. \cite{lar} for
two values of the compressibility, $K$ (in MeV);
c) Persram et al. \cite{per} for free and in-medium nucleon-nucleon 
cross sections; 
d) Gaitanos et al. \cite{gai} for static nuclear matter (LDA) and
non-equilibrium in the treatment of the collision (ACNM).
}}
\label{fig-3}
\end{figure} 

To assess the implications of our data for constraining the EoS of nuclear
matter, we compare the measurements with calculations using microscopic 
transport models. 
In Fig.~\ref{fig-3} we show the comparison of our integral $v_2$ data 
for M3 centrality with Boltzmann-Ueling-Uhlenbeck (BUU) models.
In the upper panels, the sensitivity to EoS, via the compressibility $K$, 
is investigated by comparing to calculations of Danielewicz et al. \cite{dan}
and Larionov et al. \cite{lar}
Note that, performed for $b$=5-7~fm, these calculations are for slightly
more central collisions than the data.
As both sets of calculations are for protons, together with our $Z$=1 data,
we include proton flow when available (E$_{beam}$$\ge$0.4~$A$GeV). 
As expected, the flow is significantly smaller for protons alone compared 
to $Z$=1 particles only at lower energies.
There is sensitivity of elliptic flow to EoS, but it is different as a 
function of energy in the two models: larger sensitivity is seen towards
higher energies in case of the calculations by Danielewicz et al., while
the opposite seems true in case of Larionov et al. 
Both models are close to the data, but no consistent agreement between data  
and calculations exists over all the energy range.
In particular, both models overpredict the proton data at 0.4~$A$GeV. 
As a consequence, no strong constraint on EoS can be derived at this stage.
Note that the BUU model of Danielewicz et al. was recently shown to favor 
a soft EoS at 0.4~$A$GeV, in comparison to FOPI data on the azimuthal 
dependence of kinetic energy of light particles \cite{sto}.

In the lower panels of Fig.~\ref{fig-3}, the sensitivity of elliptic flow to 
the treatment of non-equilibrium situation of the nucleus-nucleus collision 
is investigated through the calculations of Persram et al. \cite{per} 
(for all nucleons) 
and Gaitanos et al. \cite{gai} (for $Z$=1, using a coalescence approach
to produce composite particles).
Both sets of calculations are employing a soft EoS, corresponding to $K$=210
and $K$=230 MeV, respectively. 
The self-consistent treatment of in-medium nucleon-nucleon cross section 
in the model of Persram et al. \cite{per}, Fig.~\ref{fig-3}c, leads to 
a decrease of the flow magnitude and a better agreement with data,
in particular for lower energies.
The calculations of Gaitanos et al. \cite{gai}, Fig.~\ref{fig-3}d, show
that there is a large effect when including the non-equilibrium momentum 
distributions in the treatment of the collision (asymmetric colliding nuclear
matter approximation, ACNM), leading to a very good agreement with the data.
The equivalent dynamical EoS is much softer compared to static nuclear matter
under the same conditions (local density approximation, LDA). 
The difference between the two approximations is even larger than 
the span between a soft and a hard EoS has on elliptic flow.

In general, when comparing elliptic flow data and model calculations, 
a special care has to be devoted to the centrality selection, as the 
dependence of $v_2$ on impact parameter is rather strong (almost linear 
for the energies above 0.25~$A$GeV, see Fig.~\ref{fig-0}).
Another essential aspect of any comparison between data and model is the 
particle type selection \cite{and2,aaa}.
Any conclusion on EoS could be weakened by the inability of the models 
to reproduce the measured yields of composite particles.
Other observables, like directed flow \cite{and2,liu} or strange particle 
yields \cite{stu}, which may probe the collision in complementary ways, 
need to be simultaneously reproduced as well. Attempts in this direction
have already been made \cite{dan,lar,gai}.
It is unlikely that a narrow bound on EoS can be derived based on
one single observable.
Moreover, as seen in Fig.~\ref{fig-3}, the differences between various
models are still large and need to be understood.

In summary, we have presented new measurements of elliptic flow in 
Au+Au collisions for beam energies from 0.09 to 1.49~$A$GeV.
Our data span the onset of the collective expansion, through the interplay
of this expansion with the spectator shadowing, which we have studied for 
three centrality classes.
Differential elliptic flow is presented for the first time over this
broad energy range.
Interpreting the observed scaling of elliptic flow on transverse momentum
for energies above 0.4~$A$GeV as resulting from an approximate matching 
between the expansion time and the passing time of the spectators, we have 
extracted time scales for particle emission. 
Our comparison of data and model calculations showcase once more the 
difficulties that need to be solved towards extracting EoS, a goal towards
which elliptic flow is a particularly important observable. 
The precision of our $v_2$ data could in principle allow for constraining 
EoS. 
It is important that the models shall be able to reproduce all the features
of $v_2$, including its transverse momentum dependence, as well as the 
corresponding \avpt.
Further improvements and validations of current transport models through 
comparisons with experimental data are needed towards a final conclusion 
on EoS of compressed nuclear matter.

This work was partly supported by the German BMBF under Contracts No.
06HD953 and No. RUM-99/010 and by KRF under Grant No. 2002-015-CS0009.

\end{document}